\documentclass[12pt]{article}
\usepackage{amsmath,amssymb,bm,graphicx}
\usepackage{color} 
\usepackage{hyperref}
\usepackage{slashed}

\newcommand{\be}{\begin{equation}}\newcommand{\ee}{\end{equation}}
\newcommand{\bea}{\begin{eqnarray}}\newcommand{\eea}{\end{eqnarray}}
\newcommand{\brr}{\begin{array}}\newcommand{\err}{\end{array}}
\newcommand{\bit}{\begin{itemize}}\newcommand{\eit}{\end{itemize}}
\newcommand{\ben}{\begin{enumerate}}\newcommand{\een}{\end{enumerate}}

\newcommand{\bbm}{\begin{bmatrix}}\newcommand{\ebm}{\end{bmatrix}}
\newcommand{\ba}{\begin{array}}
\newcommand{\ea}{\end{array}}
\newcommand{\G}{\textbf}

\newtheorem{mydef}{Definition}
\newtheorem{Lemma}{Lemma}
\newcommand{\bd}{\begin{mydef}} \newcommand{\ed}{\end{mydef}}
\newcommand{\bthe}{\begin{theorem}} \newcommand{\ethe}{\end{theorem}}
\newcommand{\ble}{\begin{Lemma}} \newcommand{\ele}{\end{Lemma}}

\definecolor{darkred}{rgb}{.8,0,0}

\definecolor{darkblue}{rgb}{0,0,.7}

\def\intx{\int \!\!\mathrm{d}^3 {\G x}}
\def\intk{\int \!\!\mathrm{d}^3 {\G k}}

\def\lf{\left}

\def\non{\nonumber}\def\pa{\partial}\def\ran{\rangle}

\def\ri{\right}
\def\al{\alpha}\def\bt{\beta}\def\ga{\gamma}

\def\la{\lambda}\def\si{\sigma}
\def\om{\omega}

\def\1{{_{1}}}\def\2{{_{2}}}

\def\noHe0{:\;\!\!\;\!\!:H_e(0):\;\!\!\;\!\!:}
\def\noHm0{:\;\!\!\;\!\!:H_\mu(0):\;\!\!\;\!\!:}

\def\lf{\left}

\def\non{\nonumber}
\def\pa{\partial}\def\ran{\rangle}

\def\ri{\right}

\def\al{\alpha}\def\bt{\beta}\def\ga{\gamma}

\def\la{\lambda}
\def\si{\sigma}
\def\om{\omega}

\def\1{{_{1}}}\def\2{{_{2}}}

\newcommand{\tp}{\text{p}}

\usepackage{cite}

\begin{document}


\begin{center}
{\Large{\bf  Comment on the Comment on the paper\\``Can oscillating neutrino states be formulated 
universally?'' }}
\end{center}
\vskip .5 truecm
\begin{center}
{\bf {  Anca Tureanu}}
\end{center}

\begin{center}
\vspace*{0.4cm} 
{\it Department of Physics, University of Helsinki, P.O.Box 64, 
\\FIN-00014 Helsinki,
Finland
}
\end{center}
\vspace*{0.2cm} 
\begin{abstract}
Recently, our work regarding the definition of oscillating neutrino states in Eur.\ Phys.\ J.\ C {\bf 80}: 68 (2020), arxiv:1902.01232 [hep-ph] Ref. \cite{AT_neutrino} has been commented upon in arXiv:2004.04739 [hep-ph] Ref. \cite{BS}. In this note we show that, contrary to the claim in the comment, our above-mentioned work cannot be reproduced in the scheme of the so-called flavour Fock space approach described in the comment. Moreover, we prove explicitly that {\it a flavour Fock space cannot exist in a model with massive mixed neutrinos}. If the flavour Fock space scheme  were viable, it would necessarily lead, as a consequence of Coleman's theorem, to the flavour number invariance of the Hamiltonian of mixed neutrino fields, while the Hamiltonian is by construction flavour number violating. Quod est absurdum.

\end{abstract}

\section{Introduction}

The standard theoretical approach to neutrino oscillations is based on the flavour neutrino states postulated in the seminal work of Gribov and Pontecorvo in 1968 \cite{Grib_Pont}. We have recently presented a consistent and universal definition of oscillating neutrino states
as coherent superpositions of massive neutrino states \cite{AT_neutrino}. The work was motivated by the necessity to formulate within a rigorous quantum field theoretical frame the coherence of particle states of different masses, which are known to be always emitted incoherently. The idea is that in quantum field theory any state has to be created by the action of some quantum operator on the physical vacuum of the theory, in the spirit of the Klauder--Sudarshan--Glauber coherent states \cite{Klauder, Sudarshan,Glauber}. With this purpose in mind, we have been able to formulate for the first time a prescription for the definition of inherently coherent oscillating neutrino states, which are the closest possible to the  flavour neutrino states conjectured by Gribov and Pontecorvo, and with which they formally coincide in the ultrarelativistic approximation \cite{AT_neutrino}. 

In summary, the quantum field theoretical solution for describing the coherent oscillating neutrino states is to create them by the action of the creation operator for Standard Model {\it massless} neutrinos on the physical vacuum of the theory, namely the vacuum of the massive neutrino fields \cite{AT_neutrino}. In order to find such an action, we employed the method of unitarily inequivalent representations, based on Bogoliubov transformations which connect creation and annihilation operators acting in different Fock spaces (in this case, the Fock space of massless neutrinos and the Fock space of massive neutrinos).  
The method is due to Bogoliubov and it was the technical innovation which led him to the elegant explanation of superfluidity in the seminal work of 1947 \cite{Bog_superf} and superconductivity in 1958 \cite{Bog_superc}, by introducing the notion of Bogoliubov quasiparticles. At the same time, these works were signaling for the first time the concept of spontaneous breaking of symmetry. Later on, the same method was employed by Nambu and Jona-Lasinio in 1961 \cite{NJL}, in the model of dynamical generation of nucleon masses, which brought the spontaneous breaking of (chiral) symmetry in the realm of particle physics. The celebrated Haag theorem \cite{Haag} is also a product of the method of unitarily inequivalent representations. This short list of some of the most influential results in theoretical condensed matter and particle physics of the previous century shows the power and versatility of the method of unitarily inequivalent representations. Our recent works \cite{AT_neutrino, AT_neutron} testify to its potential in solving standing problems in the theory of quantum coherence, applied to the quantum field theory of particle oscillations. In this framework, the oscillating neutrino states as coherent superpositions of the massive Dirac neutrino states with equal momenta $\bf p$ and helicity $\lambda$, $|\nu_{i\lambda}({\bf p})\rangle$, $i=1,2,3$ are found to be 
\begin{eqnarray}\label{3-nu_state}
|\nu_l({\bf p},\lambda)\rangle=\sum_{i=1,2,3} U^*_{li}\sqrt{\frac{1}{2}\left(1+\frac{\tp}{\Omega_{i\tp}}\right)}|\nu_{i\lambda}({\bf p})\rangle, \ \ \ l=e,\mu,\tau,\ \ \ \tp=|{\bf p}|,\ \ \Omega_{i\tp}=\sqrt{{\tp}^2+m_i^2},
\end{eqnarray}
where $U_{li}$ are the elements of the unitary Pontecorvo--Maki--Nakagawa--Sakata mixing matrix.

The comment \cite{BS} claims that the quantization prescription proposed in \cite{AT_neutrino} can be reproduced in the so-called flavour Fock space scheme initiated in \cite{BV1}. In the following, we shall justify why the latter scheme has nothing to do with our proposal for the definition of oscillating particle states and we shall prove that a flavour Fock space for massive mixed neutrinos cannot exist.

\section{Quantum systems with interaction}\label{QFI}

To understand why the flavour Fock space for mixed neutrinos cannot exist, we shall make a short detour to the basic formulation of quantum systems with interaction, as it appears in the monographs of Bjorken and Drell \cite{BD} (pp. 90-91) and of Bogoliubov and Shirkov \cite{Bog-Shirk} (pp. 117-119). We consider relativistically invariant local quantum field theories, described by a Lagrangian density:
\be
{\cal L}(x)={\cal L}(u_i(x),\partial_\mu u_j(x)), 
\ee
where the fields $u_i(x), i=1,2,\ldots, N$ form a system with interactions. 
The system is quantized by imposing equal time commutation or anticommutation relations (ETCR), depending on the spin of the fields, just as in the free case, namely
\bea\label{etcr}
\left[u_i({\bf x},t),\frac{\partial{\cal L}}{\partial{\dot u_j}}({\bf y},t)\right]&=& i\delta_{ij}\delta({\bf x} -{\bf y}),\cr
\left[u_i({\bf x},t),u_j({\bf y},t)\right]&=& 0,\cr
\left[\frac{\partial{\cal L}}{\partial{\dot u_j}}({\bf x},t),\frac{\partial{\cal L}}{\partial{\dot u_j}}({\bf y},t)\right]&=& 0,
\eea
if we take the fields $u_i(x)$ to be of integer spin. Using Noether's theorem and the translational invariance of the Lagrangian, one obtains the Hamiltonian
\be
H=H(u, \nabla u, \dot u.)
\ee 
Knowing the Hamiltonian, one determines the equations of motion for the  quantized system.

Usually, in the case of systems with interaction, the equations of motion are not exactly solvable, therefore we cannot know their exact dependence on the space-time coordinates.
In practice, one usually starts by writting the solutions of the equations of motion at $t=0$ (i.e. in the Schr\"odinger picture) and describe their time development by the Heisenberg equations,
\be\label{H_eq}
u_i({\bf x},t)=e^{-iHt}u_i({\bf x},0)e^{iHt}.
\ee
The solution at $t=0$ is written as
\be\label{u_0}
u_i({\bf x},0)=\int\frac{d{\bf k}}{(2\pi)^{3/2}\sqrt{2\omega_{{\bf k},i}}}\big(a_{{\bf k},i}(0)e^{i{\bf k}{\bf x}}+b^\dagger_{{\bf k},i}(0)e^{-i{\bf k}{\bf x}}\big),
\ee
with $\omega_{{\bf k},i}=\sqrt{{\bf k}^2+m_i^2}$, such that, in the limit of no interaction, it coincides with the free field solution at the time $t=0$.

Imposing the ETCR \eqref{etcr} at $t=0$ and using the expansion \eqref{u_0}, one finds the 
commutation relations of $a_{{\bf k},i}(0)$ and $b_{{\bf k},i}(0)$ as for the free fields, namely:
\bea\label{cr0}
&&[a_{{\bf k},i}(0),a^\dagger_{{\bf q},j}(0)]=[b_{{\bf k},i}(0),b^\dagger_{{\bf q},j}(0)]=\delta_{ij}\delta({\bf k}-{\bf q}),\cr
&&[a_{{\bf k},i}(0),a_{{\bf q},j}(0)]=[a^\dagger_{{\bf k},i}(0),a^\dagger_{{\bf q},j}(0)]=[b_{{\bf k},i}(0),b_{{\bf q},j}(0)]=[b^\dagger_{{\bf k},i}(0),b^\dagger_{{\bf q},j}(0)]=0,
\eea
but also commutation relations between the operators and their time derivatives, for example:
\be\label{newcr0}
[a_{{\bf k},i}(0),\dot a^\dagger_{{\bf q},i}(0)]=\delta({\bf k}-{\bf q}),\ \ \ \mbox{etc}.
\ee
The exact form of the latter commutation relations depends, of course, on the coupled equations of motion satisfied by the fields $u_i(x)$.

Since the Hamiltonian is time-independent, it can be written in terms of the mode operators and their time derivatives at $t=0$,
\be\label{Ham0} H=H\Big(a_{{\bf k},i}(0), a^\dagger_{{\bf k},i}(0), b_{{\bf k},i}(0), b^\dagger_{{\bf k},i}(0),\dot a_{{\bf k},i}(0), \dot a^\dagger_{{\bf k},i}(0), \dot b_{{\bf k},i}(0), \dot b^\dagger_{{\bf k},i}(0)\Big).\ee

At this point, one can write the exact solution at an arbitrary time $t$ as
\be\label{exact_sol}
u_i(x)=u_i^+(x)+u_i^-(x)=\int\frac{d{\bf k}}{(2\pi)^{3/2}\sqrt{2\omega_{{\bf k},i}}}\big(a_{{\bf k},i}(t)e^{i{\bf k}{\bf x}}+b^\dagger_{{\bf k},i}(t)e^{-i{\bf k}{\bf x}}\big),
\ee
where 
\bea\label{H_eq_ab}
a_{{\bf k},i}(t)&=&e^{-iHt}a_{{\bf k},i}(0)e^{iHt},\ \ \ \ \ b_{{\bf k},i}(t)=e^{-iHt}b_{{\bf k},i}(0)e^{iHt},\cr
a^\dagger_{{\bf k},i}(t)&=&e^{-iHt}a^\dagger_{{\bf k},i}(0)e^{iHt},\ \ \ \ \ b^\dagger_{{\bf k},i}(t)=e^{-iHt}b^\dagger_{{\bf k},i}(0)e^{iHt}.
\eea
Using \eqref{cr0}, \eqref{newcr0} and \eqref{Ham0}, the relations \eqref{H_eq_ab} lead to some complicated nonlinear operator equations for $a_{{\bf k},i}(t), a^\dagger_i({\bf k},t)$ and $b_{{\bf k},i}(t), b^\dagger_{{\bf k},i}(t)$, which usually cannot be solved exactly. The solution of these equations is equivalent to solving the equations of motion.

The issue in which we are interested in particular is whether the operators $a_{{\bf k},i}(t)$ and $b^\dagger_{{\bf k},i}(t)$ may be regarded as annihilation and creation operators of actual particles. One has to bear in mind that the separation into positive and negative frequency parts \eqref{exact_sol} may be relativistically noninvariant, in which case the operators $a_{{\bf k},i}(t)$ and $b^\dagger_{{\bf k},i}(t)$ may not have the meaning of annihilation and creation operators \cite{Bog-Shirk, BD}. For instance, the relativistic invariance is preserved and the operators $b^\dagger_{{\bf k},i}(t)$ are creation operators only if they can be written as
\be\label{criterion1}
b^\dagger_{{\bf k},i}(t)=\sum_\alpha e^{i\omega_\alpha t} b^\dagger_{i\alpha}({\bf k}).
\ee
If, on the contrary, we have an expression with mixed positive and negative energy parts, such as
\be\label{criterion2}
b^\dagger_{{\bf k},i}(t)=\sum_\alpha e^{i\omega_\alpha t} b^\dagger_{i\alpha}({\bf k})+\sum_\beta e^{-i\omega_\beta t} b_{i\beta}({\bf k}),
\ee
the operators $b^\dagger_{{\bf k},i}(t)$  {\it can not} be interpreted as creation operators. We emphasize that in the above formulas the operators $b^\dagger_{i\alpha}({\bf k})$ and $b_{i\beta}({\bf k})$ do NOT depend on time. 

\section{Flavour Fock space does not exist}

We consider the flavour number violating Lagrangian density in the case of two-neutrino mixing with Dirac mass terms (see, for example, \cite{mohapatra,fukugita,giunti,bilenky,valle}):
\be \label{mixlag}
\mathcal{L}(x)\ =\ \overline{\nu}(x) \, \lf(i \ga^\mu \pa_\mu \ - \ M \ri) \, \nu(x) \, ,
\ee
with
\be
\nu(x) \ = \
\bbm
\nu_e (x)\\ \nu_\mu (x)
\ebm \, , \qquad
M \ = \
\bbm
m_e & m_{e \mu} \\ m_{e \mu} & m_\mu
\ebm \, .
\ee
The field equations are
\bea \label{neuteq}
\lf(i \ga^\mu \pa_\mu \ - \ M \ri)\nu(x) \ = \ 0 \, ,
\eea
i.e. the flavour fields $\nu_e (x)$ and $\nu_\mu (x)$ satisfy {\it coupled} equations of motion, namely
\bea \label{neuteq'}
(i\gamma^\mu\partial_\mu-m_e)\nu_e(x)-m_{e\mu}\nu_\mu(x)=0,\cr
(i\gamma^\mu\partial_\mu-m_\mu)\nu_\mu(x)-m_{e\mu}\nu_e(x)=0.
\eea
The Lagrangian \eqref{mixlag} is diagonalized by the unitary change of variables:
\bea
\label{PontecorvoMix}
\bbm
\nu_e (x)\\ \nu_\mu (x)
\ebm \, = \
\bbm
\cos\theta & \sin\theta\\ -\sin\theta & \cos\theta
\ebm \, \bbm
\nu_1 (x)\\ \nu_2 (x)
\ebm \, ,\eea
with $\tan 2 \theta = 2 m_{e\mu}/(m_\mu-m_e)$. The massive Dirac fields $\nu_1$ and $\nu_2$ satisfy free Dirac equations:
\bea\label{eom_m}
\lf(i \ga^\mu \pa_\mu \ - \ m_j\ri)\nu_j(x) \ = \ 0 \, , \qquad j=1,2 \, ,
\eea
where the masses $m_1$ and $m_2$ are given by the relations:
\bea
m_e & =& m_1 \, \cos^2 \theta \ + \ m_2 \, \sin^2 \theta \, , \cr
m_\mu & =& m_1 \, \sin^2 \theta \ + \ m_2 \, \cos^2 \theta \, .
\eea

According to the flavour Fock space scheme, the quantization of the theory in the set of variables $(\nu_1,\nu_2)$ leads to a Fock space of massive states, with the vacuum $|0\ran_{1,2}$, while the quantization in the variables $(\nu_e,\nu_\mu)$ would lead to an infinity of Fock space of flavour states, with the respective vacua $|0(t)\ran_{e,\mu}$ which are all orthogonal on the vacuum of massive states \cite{BHV}:
\be\label{orthog1}
_{e,\mu}\langle0 (t)|0\ran_{1,2}=0,\ \ \ \ \forall\  t,
\ee
and also orthogonal among themselves:
\be\label{orthog}
_{e,\mu}\langle0 (t)|0(t')\ran_{e,\mu}=0,\ \ \ \ \forall\  t\neq t'.
\ee
Due to \eqref{orthog}, the massive and flavour Fock spaces do not share any states, namely the flavour states cannot be written as a superposition of the massive states \`a la Pontecorvo. 

The claim of the possible existence of the flavour vacua \cite{BV1,BS} is wrong, because a unitary change of variables like \eqref{PontecorvoMix} can never modify the structure of the quantized theory.
The Fock representation is selected by the Hamiltonian and only by it \cite{SW,Strocchi}. In the following, we shall prove directly that only the vacuum  $|0\ran_{1,2}$ exists, while $|0 (t)\ran_{e,\mu}$ cannot be constructed, contrary to the claims of the flavour Fock space scheme.

\subsection{Canonical quantization of the free fields $\nu_1,\nu_2$ and their Fock space}

It is clear that the system described by the Lagrangian \eqref{mixlag} is exactly solvable by the unitary change of variables \eqref{PontecorvoMix}. This is a system of two free Dirac fields $(\nu_1,\nu_2)$ of masses $m_1,m_2$, which is easily quantized canonically.

The fields $\nu_1$ and $\nu_2$, upon quantization, are expanded as
\bea
\nu_j(x) & = & \int\frac{d{\bf k}}{(2\pi)^{3/2}\sqrt{2\omega_{{\bf k},j}}}\sum_{r} \,  \lf[u_{\mathbf{k},j}^r \, \alpha^r_{\mathbf{k},j} \, e^{-i \, \om_{\G k,j} \, t}e^{i{\bf k}\cdot {\bf x}}  + v_{\mathbf{k},j}^r \, \beta_{\mathbf{k},j}^{r\dagger} \, e^{i \om_{\G k,j} \, t} e^{-i{\bf k}\cdot {\bf x}} \ri],\ 
j=1,2 \, , \label{Fourierfield}
\eea
where the creation and annihilation operators satisfy the canonical anticommutation relations
\bea
\{\alpha^r_{\mathbf{k},j},\alpha^{\dagger s}_{\mathbf{q},j}\}=\delta_{ij}\delta_{rs}\delta({\bf k}-{\bf q}), \ \ \ \{\beta^r_{\mathbf{k},j},\beta^{\dagger s}_{\mathbf{q},j}\}=\delta_{ij}\delta_{rs}\delta({\bf k}-{\bf q}),
\eea
all the other anticommutators being zero.

The Fock space of the fields $\nu_1$ and $\nu_2$ is built on the vacuum $|0\ran_{1,2}$, which is annihilated by $\al^r_{\G k,j}$, $\bt^r_{\G k,j}$:
\bea
\al^r_{\G k,j}|0\ran_{1,2}=\bt^r_{\G k,j}|0\ran_{1,2}=0,\ \ j=1,2.
\eea

So far, we have described the standard manner of introducing Dirac masses and mixing of the different flavour fields, encountered in all the neutrino physics textbooks (see, for example, \cite{mohapatra,fukugita,giunti,bilenky,valle}). The state  $|0\ran_{1,2}$ is the physical vacuum of the theory, according to the rules of quantum field theory which state that the physical vacuum is the vacuum of the Fock space of those operators that diagonalize the Hamiltonian (see the monographs \cite{Bog-Shirk, Umezawa-book}). The vacuum $|0\ran_{1,2}$ is relativistically invariant and the lowest-lying state of the Fock space, satisfying 
\be
H|0\ran_{1,2}=0,
\ee
where 
\be\label{Ham_free}
H=\int d{\bf k}\sum_{j,r} \om_{\G k,j}(\alpha^{\dagger r}_{\mathbf{k},j}\alpha^r_{\mathbf{k},j}+\beta^{\dagger r}_{\mathbf{k},j}\beta^r_{\mathbf{k},j}).
\ee

\subsection{Quantization of the interacting fields $\nu_e,\nu_\mu$}

The fields $\nu_e,\nu_\mu$ which satisfy the coupled equations of motion \eqref{neuteq'} are regarded as {\it interacting fields}, where the interaction term in the Lagrangian is 
\be
{\cal L}_{int}(x)= -m_{e\mu}(\bar\nu_e(x)\nu_\mu(x)+\bar\nu_\mu(x)\nu_e(x)).
\ee
The parameter $m_{e\mu}$ is the "coupling constant".

We can quantize the theory as an interacting one, using the method described in Sect. \ref{QFI}. We impose the canonical equal time anticommutation relations:
\begin{eqnarray}\label{acr_nu}
\{\nu_\sigma({\bf x},t), \Pi_{\nu_{\sigma'}}({\bf y},t)\} &=&\{\nu_\sigma({\bf x},t), i\nu_{\sigma'}^{\dagger}({\bf y},t)\}=i\delta_{\sigma\sigma'}\delta({\bf x}-{\bf y}),\nonumber\\
\{\nu_\sigma({\bf x},t), \nu_{\sigma'}({\bf y},t)\} &=&0,\nonumber\\
\{\nu_\sigma^{\dagger}({\bf x},t), \nu_{\sigma'}^{\dagger}({\bf y},t)\} &=&0, \ \ \ \ \ \ \  \sigma,\sigma'=e,\mu,
\end{eqnarray}
and find the Hamiltonian corresponding to the Lagrangian \eqref{mixlag}:
\be\label{Ham_nu}
H=\int d{\bf x}\left[\sum_{\sigma=e,\mu}\left(
-\bar\nu_\sigma(x)i\gamma^{k}\partial_{k}\nu_\sigma(x) + m_\sigma\bar\nu_\sigma(x)\nu_\sigma(x)
\right)+m_{e\mu}\left(\bar\nu_e(x)\nu_\mu(x)+\bar\nu_\mu(x)\nu_e(x)\right)\right].
\ee
Using \eqref{acr_nu} and \eqref{Ham_nu} in the Hamilton equations,
\be
i\partial_t\nu_\sigma({\bf x},t)= [\nu_\sigma({\bf x},t),H],
\ee
we find the equations of motion \eqref{neuteq'}.
We write their solutions at $t=0$ as
\bea\label{sol0}
\nu_\si({\bf x},0)= \int\frac{d{\bf k}}{(2\pi)^{3/2}\sqrt{2\omega_{{\bf k},\sigma}}}\sum_{r} \, \lf[u_{\mathbf{k},\si}^r \, a^r_{\mathbf{k},\si}(0) \,  e^{i{\bf k}\cdot {\bf x}}  \ri. + \lf. v_{\mathbf{k},\si}^r \, b_{\mathbf{k},\si}^{r\dagger}(0) \,  e^{-i{\bf k}\cdot {\bf x}} \ri],\ 
\si=e,\mu \, 
\eea
where $\omega_{{\bf k},\sigma}$ as well as the spinors $u_{\mathbf{k},\si}^r, v_{\mathbf{k},\si}^r$ correspond to the masses $m_e$ and $m_\mu$, thus fulfilling the requirement that in the absence of interaction ($m_{e\mu}=0$), the solutions \eqref{sol0} coincide with the solutions of the free Dirac equation with the masses $m_e$ and $m_\mu$.

In principle, now we can continue like in Sect. \ref{QFI} and determine $H$ in terms of $a^r_{\mathbf{k},\si}(0), a^{\dagger r}_{\mathbf{k},\si}(0)$ and $b^r_{\mathbf{k},\si}(0), b^{\dagger r}_{\mathbf{k},\si}(0)$ and subsequently determine $a^r_{\mathbf{k},\si}(t), b^r_{\mathbf{k},\si}(t)$ and their hermitian conjugates by using \eqref{H_eq_ab}.

Since we know that the equations of motion \eqref{neuteq'} are diagonalized by the change of variables \eqref{PontecorvoMix}, which is valid at each and every space-time point, and the solutions for $\nu_1,\nu_2$ are already known and given by \eqref{Fourierfield}, one can write:
\bea\label{mixt0}
\nu_e({\bf x},0)&=&\cos\,\theta\nu_1({\bf x},0)+\sin\theta\,\nu_2({\bf x},0),\cr
\nu_\mu({\bf x},0)&=&-\sin\theta\,\nu_1({\bf x},0)+\cos\theta\,\nu_2({\bf x},0).
\eea
Then from \eqref{mixt0}, using \eqref{Fourierfield} and \eqref{sol0}, we find:
\bea
u_{\mathbf{k},e}^r \, a^r_{\mathbf{k},e}(0) \,   + v_{-\mathbf{k},e}^r \, b_{-\mathbf{k},e}^{r\dagger}(0) \,&=&  
\sqrt{\frac{\omega_{{\bf k},e}}{\omega_{{\bf k},1}}} c_\theta \left[u_{\mathbf{k},1}^r \, \alpha^r_{\mathbf{k},1}(0) + v_{-\mathbf{k},1}^r \, b_{-\mathbf{k},1}^{r\dagger}(0)\right]\cr
&+&\sqrt{\frac{\omega_{{\bf k},e}}{\omega_{{\bf k},2}}} s_\theta \left[u_{\mathbf{k},2}^r \, \alpha^r_{\mathbf{k},2}(0) +v_{-\mathbf{k},2}^r \, b_{-\mathbf{k},2}^{r\dagger}(0)\right], \cr
u_{\mathbf{k},\mu}^r \, a^r_{\mathbf{k},\mu}(0) \,   + v_{-\mathbf{k},\mu}^r \, b_{-\mathbf{k},\mu}^{r\dagger}(0) \, &=& -\sqrt{\frac{\omega_{{\bf k},\mu}}{\omega_{{\bf k},1}}} s_\theta\left[u_{\mathbf{k},1}^r \, \alpha^r_{\mathbf{k},1}(0) + v_{-\mathbf{k},1}^r \, b_{-\mathbf{k},1}^{r\dagger}(0)\right]\cr
&+&\sqrt{\frac{\omega_{{\bf k},\mu}}{\omega_{{\bf k},2}}} c_\theta \left[u_{\mathbf{k},2}^r \, \alpha^r_{\mathbf{k},2}(0) +c_\theta v_{-\mathbf{k},2}^r \, b_{-\mathbf{k},2}^{r\dagger}(0)\right],
\eea
where  $c_\theta\equiv \cos\theta$, $s_\theta\equiv \sin\theta$. Using for the spinors a normalization such that
\begin{eqnarray}\label{spinor_norm}
u^{r\dagger}_{{\bf k},\sigma}\,u^{r'}_{{\bf k},\sigma}&=&2\omega_{{\bf k},\sigma}\delta_{rr'},\cr
u^{r\dagger}_{{\bf k},\sigma}\,v^{r'}_{-{\bf k},\sigma  }&=&0,\ \quad \quad\quad \sigma=e,\mu,
\end{eqnarray}
we obtain
\bea \label{4x4Bog_0}  
  && \left[ \begin{tabular}{c} $a^r_{\G k,e}(0)$ \\ $b_{-\G k,e}^{r \dagger}(0)$
\\$a^r_{\G k,\mu}(0)$ \\ $b_{-\G k,\mu}^{r \dagger}(0)$ \end{tabular} \right]   = \ \left[\begin{array}{cccc}
c_\theta\, A^{\G k}_{e 1}& \, c_\theta \,B^{\G k}_{e 1} &
s_\theta \,A^{\G k}_{e 2}  &
\, s_\theta \,B^{\G k}_{e 2} \\
  \, c_\theta \,C^{\G k}_{e 1} & c_\theta\, D^{\G k}_{e 1} & \, s_\theta
\,C^{\G k}_{e 2} & s_\theta \,D^{\G k}_{e 2}
\\ 
 - s_\theta \,A^{\G k}_{\mu 1} & -\, s_\theta \,B^{\G k}_{\mu 1}& c_\theta
\,A^{\G k}_{\mu 2}
& \, c_\theta \,B^{\G k}_{\mu 2} \\ 
- \, s_\theta \,C^{\G k}_{\mu 1} & -
s_\theta\,
D^{\G k}_{\mu 1} &  \, c_\theta\, C^{\G k}_{\mu 2}& c_\theta\, D^{\G k}_{\mu 2}
\end{array}\right]
  \left[ \begin{tabular}{c} $\al^r_{\G k,1}$ \\ $\bt_{-\G k,1}^{r\dagger}$ \\  $\al^r_{\G k,2}$ \\ $\bt_{-\G k,2}^{r\dagger} $\end{tabular} \right] \, , 
\eea
with
\begin{eqnarray}\label{notation}
A^\G k_{\sigma i}  & = &\frac{1}{2\sqrt{\omega_{{\bf k},\sigma}\omega_{{\bf k},i}}}u^{r\dagger}_{{\bf k},\sigma}\,u^{r}_{{\bf k},i},
\quad\quad
B^\G k_{\sigma i}  =\frac{1}{2\sqrt{\omega_{{\bf k},\sigma}\omega_{{\bf k},i}}}u^{r\dagger}_{{\bf k},\sigma}\,v^{r}_{-{\bf k},i},\\
C^\G k_{\sigma i}  & = &\frac{1}{2\sqrt{\omega_{{\bf k},\sigma}\omega_{{\bf k},i}}}v^{r\dagger}_{-{\bf k},\sigma}\,u^{r}_{{\bf k},i},\quad\quad
D^\G k_{\sigma i}  =\frac{1}{2\sqrt{\omega_{{\bf k},\sigma}\omega_{{\bf k},i}}}v^{r\dagger}_{-{\bf k},\sigma}\,v^{r}_{-{\bf k},i}.
\end{eqnarray}
The exact expressions for the coefficients $ A^\G k_{\sigma i}, B^\G k_{\sigma i},C^\G k_{\sigma i},D^\G k_{\sigma i}$ are not important (although they can be easily calculated), it suffices to say that they are nonvanishing.
Now, using the Hamiltonian in the form \eqref{Ham_free}, we determine the time dependence of the operators (see \eqref{H_eq_ab}):
\bea \label{4x4Bog_t}  
  && \left[ \begin{tabular}{c} $a^r_{\G k,e}(t)$ \\ $b_{-\G k,e}^{r \dagger}(t)$
\\$a^r_{\G k,\mu}(t)$ \\
 $b_{-\G k,\mu}^{r \dagger}(t)$ \end{tabular} \right]   = \ \left[\begin{array}{cccc}
c_\theta\, A^{\G k}_{e 1}e^{-i \, \om_{\G k,1} \, t}& \, c_\theta \,B^{\G k}_{e 1} e^{i \, \om_{\G k,1} \, t}&
s_\theta \,A^{\G k}_{e 2}e^{-i \, \om_{\G k,2} \, t}  &
\, s_\theta \,B^{\G k}_{e 2}e^{i \, \om_{\G k,2} \, t} \\
  \, c_\theta \,C^{\G k}_{e 1}e^{-i \, \om_{\G k,1} \, t} & c_\theta\, D^{\G k}_{e 1} e^{i \, \om_{\G k,1} \, t}&  s_\theta
\,C^{\G k}_{e 2}e^{-i \, \om_{\G k,2} \, t} & s_\theta \,D^{\G k}_{e 2}e^{i \, \om_{\G k,2} \, t}
\\ 
 - s_\theta \,A^{\G k}_{\mu 1}e^{-i \, \om_{\G k,1} \, t} & -\, s_\theta \,B^{\G k}_{\mu 1}e^{i \, \om_{\G k,1} \, t}& c_\theta
\,A^{\G k}_{\mu 2}e^{-i \, \om_{\G k,2} \, t}
& \, c_\theta \,B^{\G k}_{\mu 2} e^{i \, \om_{\G k,2} \, t}\\\nonumber
 - \, s_\theta \,C^{\G k}_{\mu 1}e^{-i \, \om_{\G k,1} \, t} & -
s_\theta\,
D^{\G k}_{\mu 1} e^{i \, \om_{\G k,1} \, t}&  \, c_\theta\, C^{\G k}_{\mu 2}e^{-i \, \om_{\G k,2} \, t}& c_\theta\, D^{\G k}_{\mu 2}e^{i \, \om_{\G k,2} \, t}
\end{array}\right]
  \left[ \begin{tabular}{c} $\al^r_{\G k,1}$ \\ $\bt_{-\G k,1}^{r\dagger}$ \\  $\al^r_{\G k,2}$ \\ $\bt_{-\G k,2}^{r\dagger} $\end{tabular} \right] \, .\\
\eea
Then the solutions of the equations of motion \eqref{neuteq'}, for arbitrary time, have the form:
\bea\label{sol_t}
\nu_\si({\bf x},t)= \int\frac{d{\bf k}}{(2\pi)^{3/2}\sqrt{2\omega_{{\bf k},\sigma}}}\sum_{r} \,  \lf[u_{\mathbf{k},\si}^r \, a^r_{\mathbf{k},\si}(t) \,  e^{i{\bf k}\cdot {\bf x}}  \ri. + \lf. v_{\mathbf{k},\si}^r \, b_{\mathbf{k},\si}^{r\dagger}(t) \,  e^{-i{\bf k}\cdot {\bf x}} \ri],\ \ \ 
\si=e,\mu \, .
\eea
This completes the quantization of the interacting fields $\nu_e,\nu_\mu$, according to the rules of quantum field theory \cite{BD,Bog-Shirk}. Let us emphasize that the time dependence of the operators $a^r_{\mathbf{k},\si}(t)$ and $b_{\mathbf{k},\si}^{r\dagger}(t)$ in \eqref{sol_t} is only through phases of the type $e^{\pm i\omega_{{\bf k},i}t}$ and not $e^{\pm i\omega_{{\bf k},\sigma}t}$ (see eq. \eqref{4x4Bog_t}).

The question is whether the operators $a^r_{\mathbf{k},\si}(t)$ and $b_{\mathbf{k},\si}^{r\dagger}(t)$ with the expressions given by \eqref{4x4Bog_t} can be interpreted as annihilation and creation operators, respectively. Recalling the criteria expressed in \eqref{criterion1} and \eqref{criterion2}, the answer is definitely no, because of the presence of $e^{i \, \om_{\G k,i} \, t}$ in the expansion of $a^r_{\mathbf{k},\si}(t)$ and the presence of $e^{-i \, \om_{\G k,i} \, t}$ in the expansion of $b_{\mathbf{k},\si}^{r\dagger}(t)$. {\it Consequently, one cannot define new flavour vacua $|0 (t)\ran_{e,\mu}$ by requiring them to be annihilated by $a^r_{\mathbf{k},\si}(t)$ and $b^r_{\mathbf{k},\si}(t)$.} The operators appearing in the expansion \eqref{sol_t} do not generate flavour Fock space(s). The only vacuum that can be defined in the theory is by the quantization of the set of free massive fields $(\nu_1,\nu_2)$, namely $|0\ran_{1,2}$.

\subsection{Inconsistencies of a flavour Fock space scheme}\label{inconsist}

Despite the facts explained above, in the flavour Fock space scheme, new creations and annihilation operators were introduced. How is this possible? The answer is: by an illegitimate manipulation of formula \eqref{sol_t}, as will be shown below. 

Let us examine how the expansion \eqref{sol_t} is written in formula (10) of the comment \cite{BS}:
\bea
\nu_\si({\bf x},t) = \frac{1}{\sqrt{V}}\sum_{\mathbf{k},r} \,   e^{i{\bf k}\cdot {\bf x}} \, \lf[u_{\mathbf{k},\si}^r \, \alpha_{\mathbf{k},\si}(t) \, e^{-i \, \om_{\G k,\si} \, t} + v_{-\mathbf{k},\si}^r \, \beta_{-\mathbf{k},\si}^{r\dagger}(t) \, e^{i \om_{\G k,\si} \, t}\ri] \, ,\qquad
\si=e,\mu \, , \label{Fourierfieldf}
\eea
where $\om_{\G k,\si} = \sqrt{|\G k|^2+\mu_\si^2}$ and $\mu_\si$  are mass parameters which are (partly) specified by the requirement that in the limit when $m_{e\mu}=0$, they become identical to $m_\si$. This leads to an infinity of possibilities. (Incidentally, it was proved long ago \cite{CG} that the hypothesis that neutrinos produced or detected in charged-current weak interaction processes are described by flavor neutrino Fock states implies that measurable quantities depend on the arbitrary unphysical flavor neutrino mass parameters $\mu_\sigma$.) We shall consider in \eqref{Fourierfieldf} $\mu_e=m_e$ and $\mu_\mu=m_\mu$, in order to match the formula \eqref{sol_t} above.

The salient fact when comparing \eqref{sol_t} and \eqref{Fourierfieldf}, is that the operators $\alpha^r_{\mathbf{k},\si}(t)$ and $\beta_{-\mathbf{k},\si}^{r\dagger}(t)$ were introduced in formula (10) of \cite{BS} by hand, in order to {\it force} their interpretation as annihilation and creation operators, respectively. The only way to obtain $\alpha^r_{\mathbf{k},\si}(t)$ and $\beta_{-\mathbf{k},\si}^{r\dagger}(t)$ is by taking the operators $a^r_{\mathbf{k},\si}(t)$ and $b_{-\mathbf{k},\si}^{r\dagger}(t)$ determined by \eqref{4x4Bog_t} and multiplying them "conveniently" by a phase:
\bea\label{sol_t'}
\nu_\si({\bf x},t)= \frac{1}{\sqrt{V}}\sum_{\mathbf{k},r} \, e^{i{\bf k}\cdot {\bf x}}\Big[u_{\mathbf{k},\si}^r \, \left(a^r_{\mathbf{k},\si}(t) \,e^{i \, \om_{\G k,\si} \, t}\right)e^{-i \, \om_{\G k,\si} \, t}     +  v_{-\mathbf{k},\si}^r \, \left(b_{-\mathbf{k},\si}^{r\dagger}(t) \,  e^{-i \, \om_{\G k,\si} \, t}\right)e^{i \, \om_{\G k,\si} \, t} \Big],
\eea
such that 
\bea\label{fallacy}
\alpha^r_{\mathbf{k},\si}(t)=a^r_{\mathbf{k},\si}(t)e^{i \, \om_{\G k,\si} \, t} ,\quad\quad \beta_{-\mathbf{k},\si}^{r\dagger}(t)=b_{-\mathbf{k},\si}^{r\dagger}(t)e^{-i \, \om_{\G k,\si} \, t} .
\eea
In this way, in \eqref{Fourierfieldf} the operator $\alpha^r_{\mathbf{k},\si}(t)$ comes as if in the positive-frequency part, while $\beta_{-\mathbf{k},\si}^{r\dagger}(t)$ appears in the negative-frequency part, in the hope that this will be conducive to their interpretation as annihilation and creation operators. However, this is not the case: the so-called "creation operator" $\beta_{\mathbf{k},\si}^{r\dagger}(t)$ cannot be written as a superposition of {\it time-independent} operators, therefore, by the criterion \eqref{criterion1}, it does not have the meaning of a creation operator \cite{Bog-Shirk}. Similarly it can be argued why $\alpha^r_{\mathbf{k},\si}(t)$ cannot be an annihilation operator\footnote{As an aside, in \cite{BS} formula (11), reproduced below, the operators $\alpha^r_{\mathbf{k},\si}(t)$ and $\beta_{-\mathbf{k},\si}^{r\dagger}(t)$ are given as:
\bea \label{4x4Bog_BS}  
\left[ \begin{tabular}{c} $\alpha^r_{\G k,e}$ \\ $\beta_{-\G k,e}^{r \dagger}$
\\$\alpha^r_{\G k,\mu}$ \\ $\beta_{-\G k,\mu}^{r \dagger}$ \end{tabular} \right]  
= \ \left[\begin{array}{cccc}
c_\theta\, \rho^{\G k}_{e 1}& i \, c_\theta \,\lambda^{\G k}_{e 1} &
s_\theta \,\rho^{\G k}_{e 2}  &
i \, s_\theta \,\lambda^{\G k}_{e 2} \\
  i \, c_\theta \,\lambda^{\G k}_{e 1} & c_\theta\, \rho^{\G k}_{e 1} & i \, s_\theta
\,\lambda^{\G k}_{e 2} & s_\theta \,\rho^{\G k}_{e 2}
\\ 
 - s_\theta \,\rho^{\G k}_{\mu 1} & -i \, s_\theta \,\lambda^{\G k}_{\mu 1}& c_\theta
\,\rho^{\G k}_{\mu 2}
& i \, c_\theta \,\lambda^{\G k}_{\mu 2} \\ - i \, s_\theta \,\lambda^{\G k}_{\mu 1} & -
s_\theta\,
\rho^{\G k}_{\mu 1} &  i \, c_\theta\, \lambda^{\G k}_{\mu 2}& c_\theta\, \rho^{\G k}_{\mu 2}
\end{array}\right]
  \left[ \begin{tabular}{c} $\al^r_{\G k,1}$ \\ $\bt_{-\G k,1}^{r\dagger}$ \\  $\al^r_{\G k,2}$ \\ $\bt_{-\G k,2}^{r\dagger} $\end{tabular} \right] \, , 
\eea
where $\rho^\G k_{ab}=|\rho^\G k_{ab}|e^{i(\om_{\G k,a}-\om_{\G k,b})t}$, $\la^\G k_{ab}=|\la^\G k_{ab}|e^{i(\om_{\G k,a}+\om_{\G k,b})t}$, $c_\theta\equiv \cos\theta$, $s_\theta\equiv \sin\theta$ and
\begin{eqnarray*}
 |\rho^\G k_{a b}| & \equiv & \cos\frac{\chi_a - \chi_b}{2}, \quad
|\lambda^\G k_{a b}| \ \equiv \  \sin\frac{\chi_a -
\chi_b}{2} \, , \non \\[2mm] \label{rholambda}
\chi_a & \equiv &  \cot^{-1}\lf[\frac{k}{m_a}\ri] \, ,  \qquad m_a, \, m_b \ = \ m_1, \, m_2, \, \mu_e, \, \mu_\mu\, .
\end{eqnarray*}
The mismatch in the time dependence between \eqref{fallacy} and \eqref{4x4Bog_BS} indicates that formula (11) of \cite{BS} is incorrect.}.

Let us remark that, were the manipulation as in \eqref{sol_t'} allowed, then one could construct at will Fock spaces also for {\it any} nontrivial interacting quantum field theory, by just multiplying the equation \eqref{criterion2} by $1=e^{i\omega_{\bf k} t}e^{-i\omega_{\bf k} t}$:
\be\label{criterion2'}
b^\dagger_{{\bf k},i}(t)=e^{i\omega_{\bf k} t}\left[\sum_\alpha e^{i(\omega_\alpha-\omega_{\bf k}) t} b^\dagger_{i\alpha}({\bf k})+\sum_\beta e^{-i(\omega_\beta+\omega_{\bf k}) t} b_{i\beta}({\bf k})\right],
\ee
where $\omega_{\bf k}=\sqrt{{\bf k}^2+\mu^2}$, with $\mu$ an arbitrary mass parameter. It is well known that for nontrivial interacting quantum field theories the Fock space representation does not exist (see, for example, Refs. \cite{SW, Strocchi}).

In conclusion, the procedure employed in the flavour Fock space scheme to identify creation and annihilation operators for interacting flavour fields in \cite{BS} is deceptive and the conclusion regarding the existence of flavour vacua is wrong.

\subsubsection* {Flavour vacuum and Coleman's theorem}

Still, let us assume for a moment that the flavour vacua $|0(t)\ran_{e,\mu}$ could be defined by
\be
\al^{r}_{\G k, \si}(t) \, |0(t)\ran_{e,\mu} \ = \ \bt^{r}_{\G k, \si}(t) \, |0(t)\ran_{e,\mu} \ = \ 0 \, ,\ \ \sigma=e,\mu,
\ee
and the vacuum at $t=0$ is chosen as the physical one and denoted  $|0\ran_{e,\mu}$. We shall find some contradictions that such a scheme leads to.

Further, one can define the flavour charge operators 
by
\bea
Q_{\nu_{\si}} (t) & = &  \intx \,
:\nu_{\si}^{\dag}(x)\nu_{\si}(x):\cr
& = & \ \intk \, \lf(\al^{r\dag}_{\G k,\si}(t) \, \al^r_{\G k,\si}(t) \, - \, \bt^{r\dag}_{\G k,\si}(t) \, \bt^r_{\G k,\si}(t)  \ri) \, , \quad \si=e,\mu \, , \label{QflavLept}
\eea
where $\nu_\sigma(x)$ are given in the mode expansion \eqref{Fourierfieldf}. The vacuum is invariant under the flavour charge global transformations generated by $Q_{\nu_{\si}} (0)$:
\be\label{flavour_inv_vac}
Q_{\nu_{\si}} (0)|0\ran_{e,\mu}=0.
\ee
In the flavour Fock space scheme, the fact that the flavour states of mixed neutrinos are eigenstates of the flavour operator \eqref{QflavLept} is regarded as a merit \cite{BCJV} (see also \cite{BS}).

However, the total Hamiltonian, including the Standard Model interactions and the mixed mass terms for the neutrinos, is by construction flavour number violating \cite{mohapatra,fukugita,giunti,bilenky,valle}. 
Thus, in the flavour Fock space scheme (see \cite{BS}, Sect. 4) one has:
\begin{enumerate}
\item the physical flavour vacuum of the theory is invariant under flavour charge transformations;
\item the total Hamiltonian of the theory violates flavour symmetry.
\end{enumerate}

We assume by reductio ad absurdum that the above statements are true. Let us now recall the well-known theorem of Coleman \cite{Coleman}, which is summarized as {\it "the invariance of the vacuum is the invariance of the world"}. Coleman proved that if the vacuum is invariant under the group generated by
the space integral of the time component of a local vector current, then the Hamiltonian is invariant
also. Therefore, from the physical vacuum being flavour invariant as in \cite{BS} would follow that the Hamiltonian of the theory is also flavour invariant, which is in contradiction with the obvious fact that the Hamiltonian is flavour violating by construction. The logical conclusion is that the claim no. 1. above cannot be true, namely {\it the physical vacuum cannot be flavour invariant}.

\subsubsection*{Energy nonconservation }

The flavour vacuum is not the lowest eigenstate of the Hamiltonian:
\be
H|0(t)\ran_{e,\mu}\neq 0, \quad \forall t,
\ee
and a time dependent vacuum is not invariant under any of the external transformations involving the time, in particular under translation in time. This means that energy is not conserved in the interactions of the flavour neutrinos. (Ironically, neutrinos were introduced in the first place to save energy conservation!) 

\subsection{Oscillating neutrino states \cite{AT_neutrino} cannot be reproduced in the flavour Fock space scheme}

In the comment \cite{BS} it is claimed that the oscillating neutrino states defined in \cite{AT_neutrino} can be reproduced in the flavour Fock space scheme, by taking $\mu_e=\mu_\mu=0$ in \eqref{Fourierfieldf}. Without going into the details of the procedure employed in \cite{AT_neutrino}, we will briefly show that there is no connection with the flavour Fock space scheme.

1. As mentioned earlier in Sect. \ref{inconsist}, the only allowed values for $\mu_e,\mu_\mu$ in \eqref{Fourierfieldf} are those which become identical to $m_\si$ in the limit when $m_{e\mu}=0$. Although there is an infinity of possibilities, the choice $\mu_e=\mu_\mu=0$ is not among them, consequently, this case does not belong to the flavour Fock space scheme. As a result, eqs. (13) in the comment \cite{BS} are wrong even within that scheme.

2. In addition, the operators used as creation operators in \cite{BS}, formula (14), cannot be regarded as creation operators in any  Fock space, as they do not fulfill the general criterion \eqref{criterion1}.

3. It is a pure manipulation to use a sequence of wrong formulas in order to make the flavour Fock space scheme look formally as the states defined in \cite{AT_neutrino}.

4. The procedure developed in \cite{AT_neutrino} for defining oscillating neutrino states on the vacuum of the massive neutrinos is based on a genuine application of the method of unitarily inequivalent representations, by relating the field theory of massless Standard Model neutrinos with the field theory of massive neutrinos.
In this quantization prescription, {\it at a certain moment $t=0$ and only at that moment}, we impose the identification\footnote{Nota bene: The formulas \eqref{aa} are {\it not identical to} \eqref{mixt0}, though they look formally the same. The difference is that the fields $\psi_{\nu_e},\psi_{\nu_\mu}$ are solutions of the massless Dirac equations \eqref{aaaa}, and not of the coupled equations \eqref{neuteq'}.}
\bea
\label{aa}&&\nu_1 ({\bf x},0) = \psi_{\nu_e} ({\bf x},0)\cos\theta - \psi_{\nu_\mu}({\bf x},0)\sin\theta \, , \cr
&&\nu_{2}({\bf x},0) = \psi_{\nu_e}({\bf x},0)\sin\theta + \psi_{\nu_\mu} ({\bf x},0)\cos\theta \, , 
\eea
when the fields $\nu_1,\nu_2$ and $\psi_{\nu_e},\psi_{\nu_\mu}$ are solutions of the equations of motion
\bea\label{aaa}
&&\lf(i \ga^\mu \pa_\mu \ - \ m_j\ri)\nu_j(x) \ = \ 0 \, , \qquad j=1,2 \, ,\\\label{aaa}
&&i \ga^\mu \pa_\mu \ \psi_{\nu_\sigma}(x) \ = \ 0 \, , \qquad \sigma=e,\mu \, .\label{aaaa}
\eea
The set of equations \eqref{aa}, \eqref{aaa} and \eqref{aaaa} {\it is compatible}, as it means that the solution of a massive Dirac equation and the solution of a massless Dirac equation coincide at a given time moment, but otherwise evolve according to their respective Hamiltonians. Such a set of equations is allowed by the method of unitarily inequivalent representations and similar equations are encountered in the papers of Nambu and Jona-Lasinio \cite{NJL}  (see also \cite{UTK, Umezawa-book}) or Haag \cite{Haag}. The reasons behind the equations \eqref{aa} and \eqref{aaa} are amply explained in Ref. \cite{AT_neutrino} and we feel that it is unnecessary to repeat here those detailed explanations. It suffices to say that \eqref{aa} and \eqref{aaa} are required in order to diagonalize the Hamiltonian corresponding to the flavour number violating Lagrangian \eqref{mixlag}, and to establish the Bogoliubov transformations between the creation and annihilation operators of the massive neutrino fields (corresponding to the observable quasiparticles) and the operators of the massless flavour neutrino fields (corresponding to the inobservable bare particles). In the end, the coherent oscillating particle states are defined by the application of the massless neutrino creation operators to the vacuum of the massive neutrinos, and their exact form is given in eq. \eqref{3-nu_state}.

\section{Conclusions}

We have proven explicitly that {\it flavour neutrino Fock spaces cannot be constructed}, when the flavour neutrino fields are linear combinations of massive neutrino fields with different masses  as in \eqref{PontecorvoMix}. 
Indeed, the equations of motion for the free fields $(\nu_1,\nu_2)$ are generated by the same Hamiltonian as the equations of motion for the interacting fields $(\nu_e,\nu_\mu)$. Thus, the two sets of fields are related by a unitary change of variable, therefore they represent the same degrees of freedom.  The vacuum of the theory described by the Lagrangian \eqref{mixlag} is unique. This already invalidates the claims made in the comment \cite{BS}.

Furthermore, the flavour Fock space scheme cannot be viable by very general arguments, such as Coleman's theorem \cite{Coleman}, since it leads to the absurd conclusion that the Hamiltonian of mixed neutrino fields is, simultaneously, flavour number violating and flavour number symmetric.

Our work on the formulation of oscillating neutrino states \cite{AT_neutrino} is completely disconnected from the so-called flavour Fock space scheme described in the comment \cite{BS} and in the references therein.

The assumption of the existence of time-dependent flavour vacua is based on a fallacious definition of creation and annihilation operators in interacting theories \eqref{fallacy}, in contradiction with the principles of quantum field theory \cite{BD,Bog-Shirk, SW, Strocchi}. Additionally, such an assumption would lead to sensational consequences, such as the 	energy nonconservation in interactions, along with the breaking of several other symmetries. It would have been really a great surprise if the trivial system of two non-interacting massive Dirac fields were to hide inside so much drama.

\end{document}